\documentclass[twocolumn]{aastex63}
\title{2I-CN}

\newcommand\aastex{AAS\TeX}

\usepackage{xcolor}
\usepackage{url}

\received{September 26, 2019}
\revised{September 30, 2019}
\accepted{October 1, 2019}

\submitjournal{ApJ Letters}

\shorttitle{\aastex\ Detection of CN gas in 2I/Borisov}
\shortauthors{Fitzsimmons et al.}

\begin{document}

\title{Detection of CN gas in Interstellar Object 2I/Borisov}

\correspondingauthor{Alan Fitzsimmons}
\email{a.fitzsimmons@qub.ac.uk}

\author[0000-0003-0250-9911]{Alan Fitzsimmons}
\affiliation{Astrophysics Research Centre \\
Queen's University Belfast \\
Belfast BT7 1NN, UK}

\author[0000-0001-6952-9349]{Olivier Hainaut}
\affiliation{European Southern Observatory \\
Karl-Schwarzschild-Strasse 2\\
D-85748 Garching bei M\"unchen, Germany}

\author[0000-0002-2058-5670]{Karen J. Meech}
\affiliation{Institute for Astronomy \\
2680 Woodlawn Drive \\
Honolulu, HI 96822 USA}

\author{Emmanuel Jehin}
\affiliation{STAR Institute \\
Universit\'e de Li\`ege\\
All\'ee du 6 ao\^ut, 19C\\
4000 Liege, Belgium}

\author[0000-0001-9784-6886]{Youssef Moulane}
\affiliation{European Southern Observatory \\
Alonso de Cordova 3107\\
Vitacura, Santiago, Chile}
\affiliation{STAR Institute \\
Universit\'e de Liege\\
All\'ee du 6 ao\^ut, 19C\\
4000 Liege, Belgium}
\affiliation{Oukaimeden Observatory\\
Cadi Ayyad University, Morocco}

\author[0000-0002-9298-7484]{Cyrielle Opitom}
\affiliation{European Southern Observatory \\
Alonso de Cordova 3107\\
Vitacura, Santiago, Chile}

\author{Bin Yang}
\affiliation{European Southern Observatory \\
Alonso de Cordova 3107\\
Vitacura, Santiago, Chile}

\author[0000-0002-2021-1863]{Jacqueline V. Keane}
\affiliation{Institute for Astronomy \\
2680 Woodlawn Drive \\
Honolulu, HI 96822 USA}

\author[0000-0002-4734-8878]{Jan T. Kleyna}
\affiliation{Institute for Astronomy \\
2680 Woodlawn Drive \\
Honolulu, HI 96822 USA}

\author[0000-0001-7895-8209]{Marco Micheli}
\affiliation{ESA NEO Coordination Centre \\
Largo Galileo Galilei, 1 \\
00044 Frascati (RM), Italy}
\affiliation{INAF - Osservatorio Astronomico di Roma \\
Via Frascati, 33 \\
00040 Monte Porzio Catone (RM), Italy}

\author[0000-0001-9328-2905]{Colin Snodgrass}
\affiliation{Institute for Astronomy,\\  
University of Edinburgh, \\
Royal Observatory, \\
Edinburgh EH9 3HJ, UK }

\begin{abstract}
The detection of Interstellar Objects passing through the Solar System offers the promise of constraining the physical and chemical processes involved in planetary formation in other extrasolar systems. While the effect of outgassing by 1I/2017 U1 ('Oumuamua) was dynamically observed, no direct detection of the ejected material was made. The discovery of the active interstellar comet 2I/Borisov means spectroscopic investigations of the sublimated ices is possible for this object.
We report the first detection of gas emitted by an interstellar comet via  the near-UV emission of CN from 2I/Borisov at a heliocentric distance of $r$ = 2.7 au on 2019 September 20. The production rate was found to be Q(CN) = $(3.7\pm0.4)\times10^{24}$ s$^{-1}$, using a simple Haser model with an outflow velocity of 0.5 km s$^{-1}$. No other emission was detected, with an upper limit to the production rate of C$_2$ of $4\times10^{24}$ s$^{-1}$. The spectral reflectance slope of the dust coma over $3900$ \AA\ $< \lambda< 6000$ \AA \ is steeper than at longer wavelengths, as found for other comets. Broad band $R_c$ photometry on 2019 September 19 gave a dust production rate of $Af\rho=143\pm10$ cm. Modelling of the observed gas and dust production rates constrains the nuclear radius to $0.7-3.3$ km assuming reasonable nuclear properties. Overall, we find the gas, dust and nuclear properties for the first active Interstellar Object are similar to normal Solar System comets.

\end{abstract}



\section{Introduction} 
\label{sec:intro}

Solar System formation models suggest that a large number of planetesimals were ejected to space as the giant planets formed and migrated. Most of these planetesimals are expected to be icy (i.e. comet-like) with only a small fraction of them being rocky objects \citep{meech2016,engelhardt2017}. Assuming that similar processes have taken place elsewhere in the Galaxy, a large number of planetesimals are wandering through interstellar space, some eventually crossing the Solar System. Many decades of comet and asteroid studies have yielded formation models that explain the mass distribution, chemical abundances, and planetary configuration of the Solar System today. However, studies of exoplanet systems  have shown that many different planetary system architectures can exist.
It is still uncertain whether the Solar System is typical of planetary systems in general. Interstellar Objects (ISOs) provide an opportunity to study the planet-building process in extrasolar planetary systems.

The first known ISO, 1I/2017 U1 ('Oumuamua), was discovered on 2017 October 19 and followed by a short intense period of observation as it faded quickly as it receded from Earth. 
Assuming it was dark, 1I was very red and small with an average diameter of 200m \citep{meech2017}. However, it could possibly be as small as 100 m across if it has a higher albedo \citep{trilling2018}.
 `Oumuamua's rotational light curve was extraordinary, with a brightness range of over 2.5 magnitudes, implying that it had a very elongated nucleus with an axis ratio $>$5:1, perhaps as large as 10:1.  Further, it was found to be in an excited rotation state with a period of 8.67$\pm$0.34 h precessing around the angular momentum vector, and a longer period of $\sim$54 h.  Because of the long damping timescale, this excited state was likely caused by its ejection from its home star system \citep{belton2018,fraser2018,drahus2018}.  Although there were very sensitive searches for dust and gas \citep{meech2017,ye2017,trilling2018}, none were detected.
Spectroscopy only revealed a red featureless spectrum similar to that expected for an irradiated cometary surface \citep{fitzsimmons2018}.
Nevertheless, non-gravitational accelerations were detected in 1I's motion combining astrometry from the ground and with the long extension of arc afforded by HST observations \citep{micheli2018}.  The only plausible explanation for this acceleration was comet outgassing at a level below our ability to detect it during the $\sim$2-week period that we had for detailed observations.  

One of the most important questions left unanswered from the study of 'Oumuamua was ``What is it made of?'' It had been expected that any ISO would probably be ice-rich, displaying cometary activity if it passed within the sublimation distance of the Sun for its constituent ices. This in turn would allow spectroscopy of the coma gases. The near-inert nature of 1I was initially puzzling, although subsequent studies have shown it is consistent with ejection mechanisms and subsequent evolution in interstellar space (see \cite{issi2019} and references within).

On 2019 August 30 Gennady Borisov at the MARGO observatory in Crimea discovered Comet C/2019 Q4 (Borisov) at small solar elongation in the morning twilight. The orbit was very quickly shown to be hyperbolic, with an eccentricity $>$ 3 (MPEC 2019-R106; 2019 September 11). On September 24 it was officially named by the IAU as 2I/Borisov - the second known ISO. Unlike 1I, this second ISO was discovered before perihelion ($q$ = 2.0 au; 2019 December 8) and will be well placed for observing before it goes into solar conjunction again in 2020 October. Initial photometry showed it to possess broad-band optical colors similar to other active comets with significant dust comae \citep{guzik2019}. An optical spectrum showed a featureless red reflectance spectrum \citep{de_Leon_2019}. However, to probe the composition of an ISO and compare its nature to our own Solar System requires identification and measurement of emission or absorption features within its spectrum. In this paper we report the first detection of gas in the coma of an interstellar comet.
 
\section{Observations and Data Reduction} 
\label{sec:observations}

2I/Borisov was observed with the 4.2 m William Herschel Telescope (hereafter WHT) plus ISIS spectrograph on La Palma on 2019 September 20.2 UT. The observational circumstances are given in Table~\ref{tab-obslog}.
The ISIS R300B grating was used with an intrinsic spectral resolving power of $\lambda/\delta\lambda=976$.
The detector was  a blue-sensitive EEV-4280 CCD, giving a pixel scale of 0.86 \AA \ pixel$^{-1}$ at 4000 \AA . Two 900-second and two 1200-second exposures
were obtained through a 2\arcsec\ wide slit, chosen to maximise the cometary flux while minimising the background flux due to airglow, astronomical twilight and moonlight.
This slit width decreased the effective resolution to 9.5 \AA\ at 4000\AA.
Due to the faintness of the comet, atmospheric extinction
at  airmasses $\geq 2.0$
and the rapidly brightening sky, only two of the four exposures obtained were found to contain cometary flux at wavelengths $<4500$ \AA , 
starting at 05:19 UT (900 second exposure) and 05:38 UT (1200 second exposure).
An exposure of the spectrophotometric standard G191-B2B \citep{bohlin1995} was obtained immediately afterwards using a 10\arcsec\ wide slit to enable flux calibration.

 \begin{deluxetable*}{lcccccccc}
 \tablecaption{Log of Photometric (CFHT, TN) and spectroscopic (WHT) observations of 2I/Borisov. CFHT magnitudes were measured through a 5\arcsec aperture, TN through a 4.2\arcsec aperture.} \label{tab-obslog}
 \tablecolumns{11}
 \tablehead{
 \colhead{Date}& 
 \colhead{Telescope} & 
 \colhead{r$_H$} &
 \colhead{Delta} &
 \colhead{$r_{PS1}$} & 
 \colhead{$B$} &
 \colhead{$V$} &
 \colhead{$R_c$} &
 \colhead{$I_c$} \\
 \colhead{ (UT)} & \colhead{} & \multicolumn{2}{c}{(au)} & \colhead{} & \colhead{} & \colhead{} & \colhead{} &  \colhead{} 
 }
 \startdata
 \hline
 Sep 9.6 & CFHT  & 2.81 & 3.48 & $18.00\pm0.03$ & & & & \\
 Sep 10.6 & CFHT & 2.80 & 3.45 & $18.00\pm0.03$ & & & & \\
 Sep 11.2 & TN &  2.79 & 3.44 & -- & 19.04$\pm$0.11 &18.11$\pm$0.06 &17.68$\pm$0.05 &17.08$\pm$0.06  \\
 Sep 12.2 & TN &  2.77 & 3.42 & -- &19.06$\pm$0.12 &18.19$\pm$0.07 &17.83$\pm$0.05 &17.29$\pm$0.08 \\
 Sep 16.2 & TN & 2.71 & 3.33 & -- & -- &18.00$\pm$0.04 &17.59$\pm$0.03 &17.25$\pm$0.03 \\
 Sep 19.2 & TN &  2.67 & 3.27 & -- & 18.88$\pm$0.11 &17.92$\pm$0.04 &17.45$\pm$0.04 &17.05$\pm$0.05 \\
 Sep 20.2 & WHT & 2.66 & 3.24 & -- & -- & -- & -- & -- \\
 \hline
 \enddata
 \end{deluxetable*}

The WHT spectra were bias subtracted and flatfielded. The wavelength calibration used CuNe+CuAR exposures made directly after the observations of the comet, resulting in an rms uncertainty of $0.04 $\AA \ in the range $3200 - 6000$ \AA . The spectrum of the comet was extracted over 8\arcsec\ centered on the comet, with the background sky measured 10\arcsec\ to 30\arcsec\ from the nucleus. Flux calibration was performed using the G191-B2B spectrum assuming the standard atmospheric extinction curve for La Palma \citep{king1985}. There was thin cloud present during the observations, and combined with the high airmass, we caution that the uncertainty in the flux from the comet could potentially be at the level of tens of percent.

As part of our campaign to get astrometric observations of 2I to confirm that its orbit was hyperbolic, we also obtained images using the Canada-France-Hawaii Telescope (hereafter CFHT) and MegaCam on 2019 Sep. 9 and 10. Megacam covers a $1\times1$ square degree field of view at a pixel scale of $0.18$\arcsec/pixel. The data were obtained through an SDSS $r'$-band filter. Our pipeline processing performs bias subtraction and flatfielding, and calibrates images against the Pan-STARRS DR2 database \citep{flewelling2016} to provide a photometric zero point for each frame.

Additionally, $BV R_c I_c$ photometry of 2I was performed with TRAPPIST-North (hereafter TN) located at Oukaimeden observatory, Morocco \citep{jehin2011}.  TN is equipped with a 2K$\times$2K  CCD camera with a field of view of 22\arcmin$ \times$22\arcmin , the pixels are binned $2\times$ by 2 to give a plate scale of 1.2\arcsec/pixel.  
Attempts were made on other dates but were thwarted by moonlight. With TN there was only a 30 minute window to observe the comet  $>25^\circ$ above the horizon. Data calibration followed standard procedures using frequently updated master bias, flat and dark frames. The removal of the sky contamination and the flux calibration were performed using TN zero points that are regularly updated.
Observational circumstances of all photmetry are given in Table~\ref{tab-obslog}.

\begin{figure*}[ht!]
\begin{center}
\includegraphics[scale=0.75]{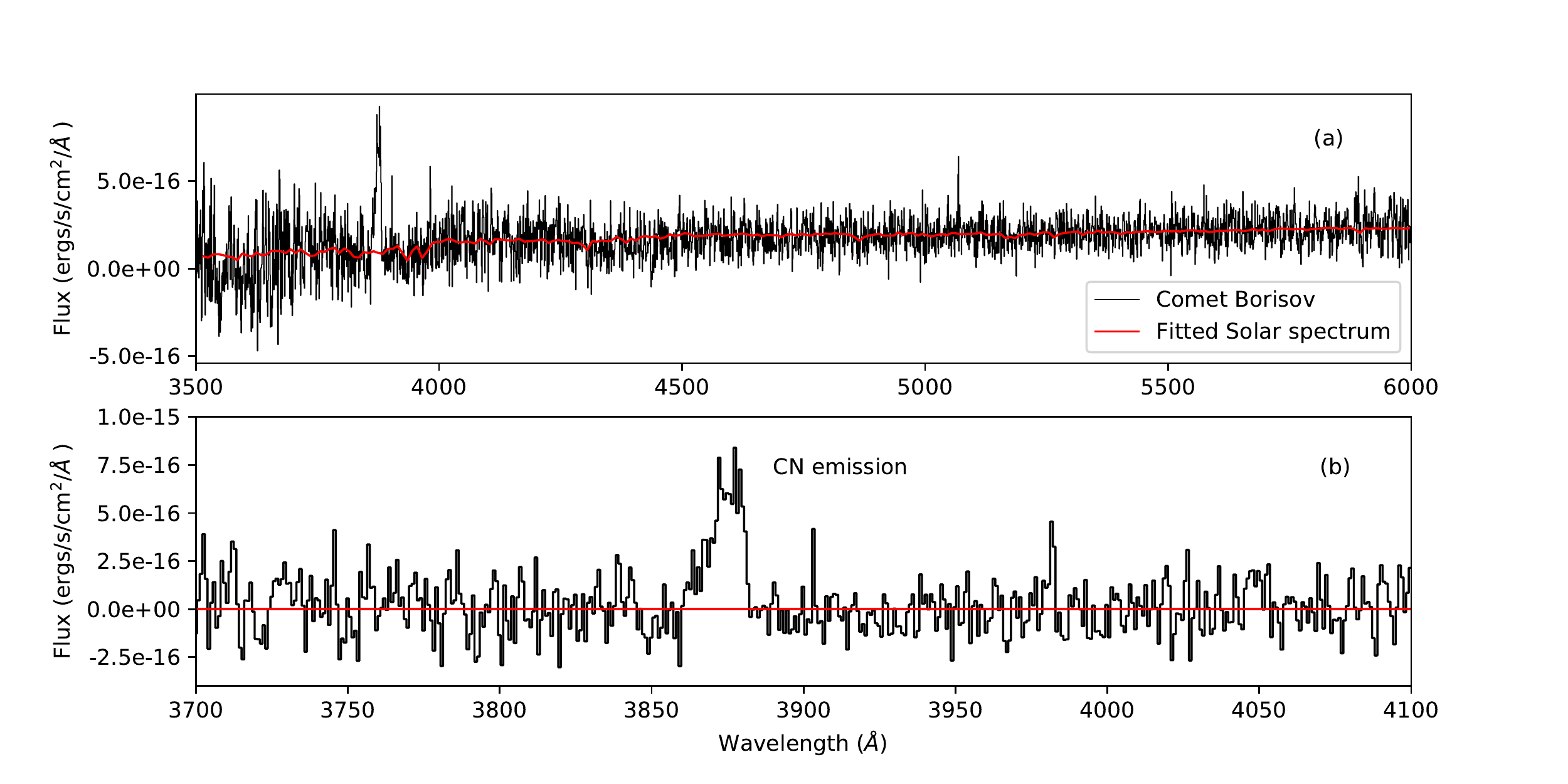}
\caption{(a) Flux calibrated spectrum of 2I/Borisov through a 2\arcsec\ by 8\arcsec\ aperture centered on the comet. Also shown is a scaled solar spectrum reddened to match the observed dust continuum. (b) Spectral region around the CN (0-0) emission band with the background dust continuum subtracted.}
\label{fig:CN}
\end{center}
\end{figure*}

\section{Analysis}
\subsection{Gas Emission}
\label{sec:analysis}
To identify any emission bands, we removed the underlying continuum resulting from reflected sunlight by dust. We used a scaled and reddened standard reference solar spectrum. After subtracting this spectrum, the comet spectra should only consist of gas fluorescence emission features. The spectrum before and after removal of the dust continuum is shown in Fig.~\ref{fig:CN} and shows strong CN (0-0) gas emission due to solar fluorescence at 3880 \AA . Both the wavelength and the asymmetric profile confirm this as CN. 

The CN(0-0) emission was seen in both analysed spectra, with an equal intensity within the measurement uncertainties. Unfortunately it was found that combining these spectra resulted in slightly lower S/N than using only the spectrum obtain at 05:38 UT. This was due to the 05:19 UT spectrum being at higher airmass and with higher sky background. Hence we only used the 05:38 UT spectrum for analysis. The CN emission was directly measured by approximating the band shape with two free-fitted gaussians, giving a flux of $(8.4\pm0.9)\times10^{-15}$ ergs s$^{-1}$ cm$^{-2}$. Using the fluorescence scattering efficiency factors from \cite{schleicher2010}, this gives $(1.2\pm0.1)\times10^{26}$ CN molecules within the extraction aperture. 

We used a simple optically-thin Haser model \citep{haser1957} to calculate gas production rates using the scalelengths from \cite{ahearn1995}. We have adopted an outflow velocity of $0.85/\sqrt{r_h} = 0.5$ km s$^{-1}$ \citep{cochran1993}. Due to the small size of the spectroscopic aperture, we numerically integrated the Haser model within the slit to derive the corresponding production rate of CN. We find Q(CN)$=(3.7\pm 0.4)\times 10^{24}$ s$^{-1}$. Assuming this gas is only created through the photodissociation of HCN, this would be the sublimation rate of this parent molecule from the nucleus.

\begin{figure*}[ht!]
\begin{center}
\includegraphics[scale=0.5]{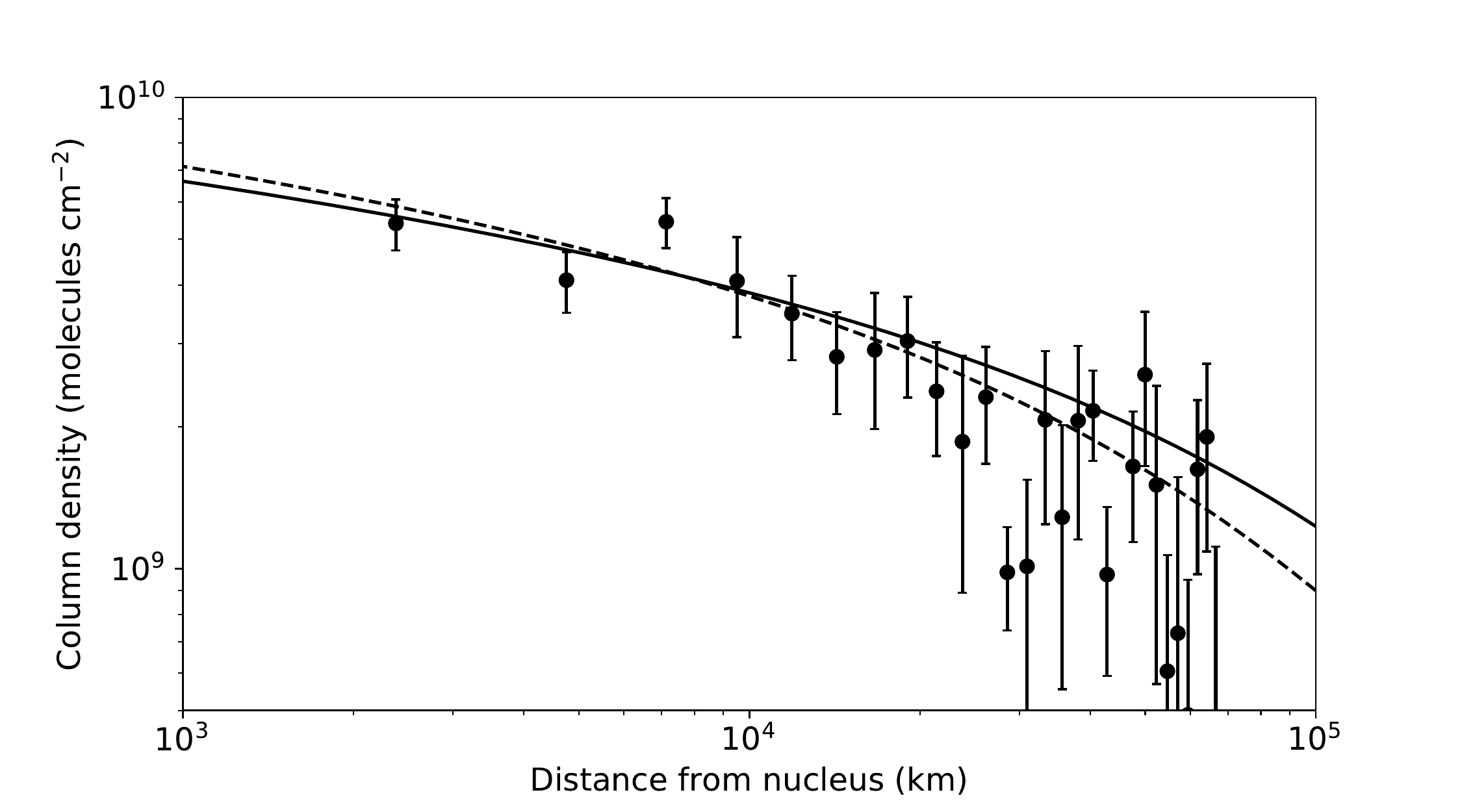}
\caption{CN column density as a function of distance from the nucleus of 2I/Borisov. The solid line indicates the Haser model assuming scalelengths that only vary as $r_h^2$. The dashed line shows the same model but with a parent scale length that scales as outflow velocity $v$.}
\label{fig:CNprofile}
\end{center}
\end{figure*}

Fig.~\ref{fig:CNprofile} shows the spatial profile of the CN column density in our data.
Overlaid is a
predicted
Haser column density profile assuming the same A'Hearn et al. scalelengths. We note that there is some indication of a faster fall-off in column density than normally observed. However, these data were dominated by the bright background sky where small changes in the fitted sky background can give rise to large changes in the measured flux. There is also the aspect that the nominal scalelengths from A'Hearn et al. are only scaled as heliocentric distance $r_h^2$, assuming an outflow velocity of 1 km $^{-1}$. While the Haser model is not physically realistic, we can approximate the effect of our slower assumed outflow velocity by scaling the parent scalelength (for HCN as the parent molecule, the daughter velocity will be dominated by the photodissociation energy and hence will be less affected). The resulting Haser profile can be seen in Fig.~\ref{fig:CNprofile} to provide a better match to the measured column densities. Therefore we conclude that within measurement and modelling uncertainties, there is no significant evidence for different CN scalelengths in these data compared to other comets.

No other emission features were apparent in our spectra.
In active Solar System comets, the second most prominent emission feature in optical spectra is the C$_2$(0-0) emission band with a band head at 5167\AA . Around this wavelength our WHT spectrum has an rms uncertainty of $8.1\times10^{-17}$ ergs s$^{-1}$ cm$^{-2}$ \AA$^{-1}$. Using an effective bandwidth of $\sim 100$\AA\ for this emission band, and using the standard relationship for spectroscopic upper limits \citep{cochran2012}, we find a $3\sigma$ upper limit to the C$_2$(0-0) flux of $6\times10^{-15}$ ergs s$^{-1}$ cm$^{-2}$. Performing a similar analysis to CN with Haser model scalelengths from \cite{ahearn1995}, we derive an upper limit of Q(C$_2$)$\leq4\times 10^{24}$ s$^{-1}$. 

\subsection{Dust Continuum}
The lack of observable gas emission at $\lambda >3900$ \AA \ allows a clean measurement of the coma dust reflectance spectrum (assuming a negligible contribution from the nucleus). We divided the fluxed comet spectrum by the standard solar spectrum and normalised the calculated reflectance spectrum to 1 at 5500 \AA . This resulted in a linear spectrum in the range $3900$ \AA$<\lambda<6000$ \AA \ with a slope of $19.9\pm 1.5$\%/$10^3$ \AA . This is approximately twice as steep as that reported by \cite{de_Leon_2019} in the range $5500$ \AA$<\lambda<9000$ \AA . However, we note that their published dust reflectance spectrum appears to become steeper at $\lambda<6000$ \AA, where our data lie. In support of this, the mean colors of the comet measured from our TN imaging data through a  4.2\arcsec\ aperture (corresponding to about a 10000\,km radius) over the preceeding 9 days were $(B-V)=0.92\pm0.06$ and $(V-R_c)=0.41\pm0.01$. These correspond to spectral slopes of 24\%/$10^3$ \AA\ over
$4380$ \AA$ <\lambda<5450$ \AA\ and 6\%/$10^3$ \AA\ over $5450$ \AA $<\lambda<6410$ \AA . All these data agree with the general spectral behavior of normal comet dust being redder at shorter wavelengths \citep{jewitt1986}.
We conclude that our data are consistent in finding a steep spectral reflectance slope at blue-visual wavelengths. 

As the WHT spectra showed no gas emission in the $B$ and $V$ bands, 
we have used the TN
magnitudes
to calculate relative dust production rates using the $Af\rho$ formalism of \cite{ahearn1984}
and correcting to zero degrees phase angle using the composite dust phase function of D. Schleicher\footnote{https://asteroid.lowell.edu/comet/dustphase.html}.
We find relative dust production rates for a radius of $10^4$ km in each filter of $(Af\rho)_B=88\pm15$ cm, $(Af\rho)_V=140\pm15$ cm,   $(Af\rho)_{R_c}=143\pm10$ cm and $(Af\rho)_{I_c}=142\pm13$ cm. Although these data were obtained 24 hours before the WHT spectra, the TN monitoring shows no significant evolution of the coma brightness over the previous 9 nights, and so we take these values as representative of the coma on September 20.2 UT.

\section{Discussion}
\label{sec:discuss}

\subsection{Comparison with other active comets}

2I/Borisov at 2.7 au appears to be similar but slightly less active than many long-period comets observed at similar distances. These include C/2013 R1 (Lovejoy) with Q(CN)$=1.9\times10^{25}$  s$^{-1}$ at 2.7 au post perihelion \citep{opitom2015}; C/2013 A1 (Siding Spring) with Q(CN)$=1.1 \times 10^{25}$  s$^{-1}$ at 2.4 au pre-perihelion \citep{opitom2016};  C/2014 W2 (PANSTARRS) with Q(CN)$=5.3\times 10^{24}$  s$^{-1}$ at perihelion at 2.7 au \citep{hyland2019}. Short-period comets often display weaker outgassing rates. 9P (Tempel 1) was found to have Q(CN) =$1.8 \times 10^{23}$ s$^{-1}$  inbound at 2.4 au \citep{meech2011}, while 67P (Churyumov-Gerasimenko) had Q(CN)$=1.3 \times 10^{24}$  s$^{-1}$  inbound at 1.3 au and Q(CN)$=9.0 \times 10^{23}$  s$^{-1}$ outbound at 2.9 au \citep{opitom2017}. Hence it is clear that the range of gas production rates measured for Solar System comets spans our measurement of 2I/Borisov.

\begin{figure*}[ht!]
\begin{center}
\includegraphics[scale=0.45]{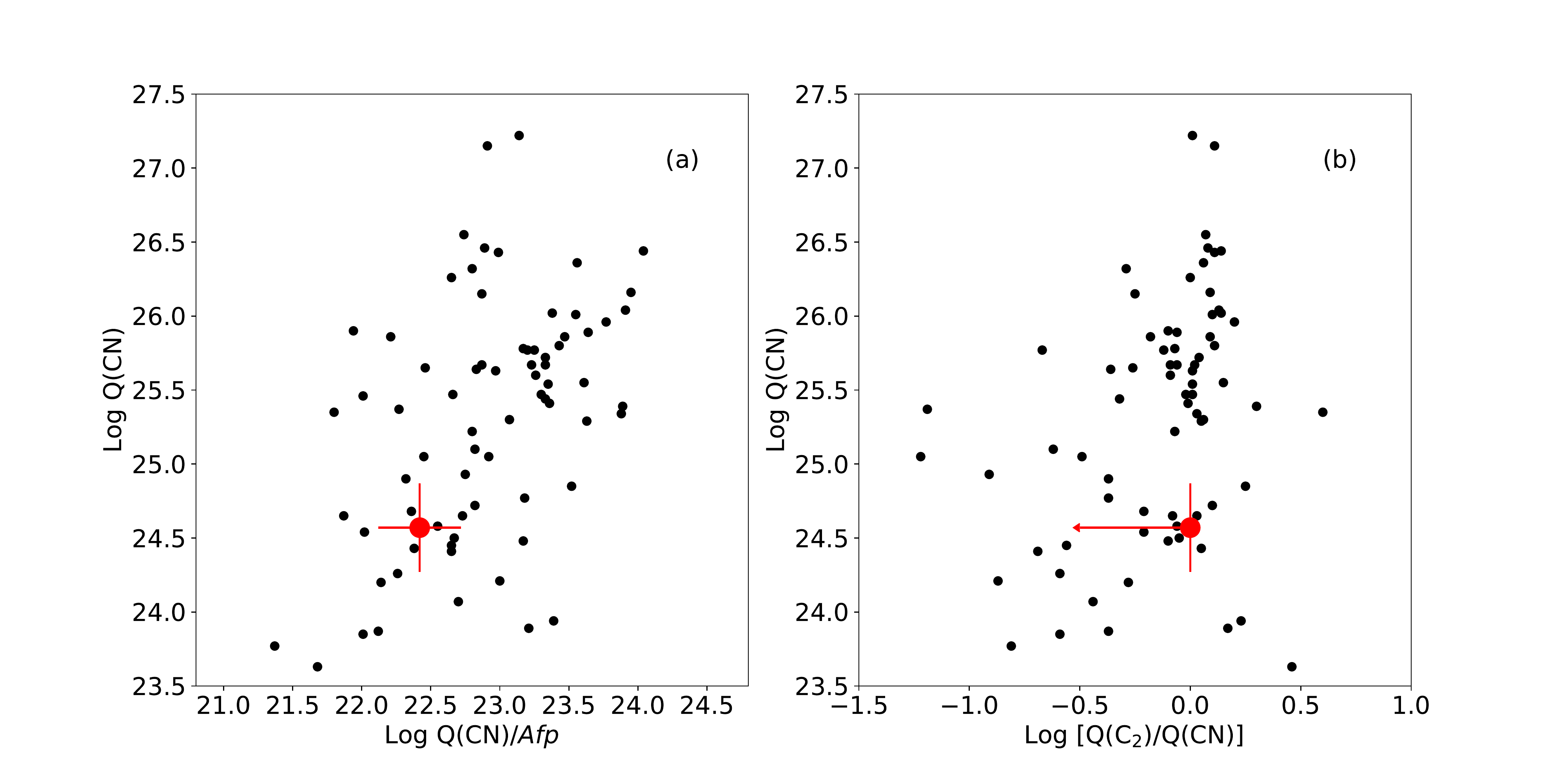}
\caption{(a) Comparison between 2I (red data point) and Q(CN)/$Af\rho$ from the \cite{osip2003} database for a variety of comets with different Q(CN). We use $(Af\rho)_V$ for 2I to match the green continuum data from the Lowell database. Uncertainties on  2I assume a conservative 50\% uncertainty due to thin cloud at the time of observation. Error bars for the Lowell database are omitted for clarity.  (b) As for (a) but for Q(C$_2$)/Q(CN) indicating our upper limit for Q(C$_2$).}
\label{fig:CNabundances}
\end{center}
\end{figure*}

Similarly, we find that the gas/dust and relative gas production rates are not obviously different to `normal' Solar System comets. In Fig.~\ref{fig:CNabundances}  we compare the large number of  measurements of Q(CN)/Q(C$_2$) and Q(CN)/$Af\rho$ from the Lowell Observatory Comet Photometry Database \citep{osip2003} which used the same Haser model scalelengths. It should be noted that these comets were a mixture of Long-Period and Short-Period Comets observed over a range of heliocentric distances. Also, the Lowell database production rates assume $v=1$ km s$^{-1}$ for all comets; using this for our data would decrease Q(CN) and our upper limit to Q(C$_2$) by a factor 2. Nevertheless, it is apparent from these data that both the gas/dust and the relative amounts of CN and C$_2$ produced by 2I/Borisov are consistent with the bulk population of Solar System comets previously measured.

Given this similarity, we can use the above measurements to estimate the possible gas-to-dust-ratio emitted by the nucleus, {\em assuming} the measured properties of Solar System comets. For the gas mass loss rate, this is straightforward. Normal comets have Q(H$_2$O)/Q(HCN)$\simeq500$ \citep{Bockleemorvan2004}, 
which for our derived Q(CN) would imply  Q(H$_2$O) $=1.7 \times10^{27}$ s$^{-1}$, assuming all CN results from the dissociation of HCN. Assuming the gas mixing ratio, 77\% water, 13\% CO and 10\% other gas, then the gas mass loss rate is $dM_g\simeq3.4\times10^{-26}\times$Q(H$_2$O)$\simeq57$ kg s$^{-1}$.

It is more difficult to estimate the dust production rate using $Af\rho$. 
Using the observed $R_c$ magnitude and the simplistic  assumptions of a single dust grain diameter of 1 $\mu$m, a grain albedo of $0.04$, density of $1000$ kg m$^{-3}$, and radially outflowing dust with velocity  $V_d=100$ m s$^{-1}$, then Q(dust)$\simeq1$ kg s$^{-1}$. While this implies a very low dust/gas ratio, some Solar System comets such as 2P/Encke have this characteristic. On the other hand, assuming larger dust particles of size 20 $\mu$m (see below) results in Q(dust)$\simeq30$ kg s$^{-1}$ and a dust/gas ratio $\sim 1$. This is closer to, but still lower than, in-situ {\it Rosetta} pre-perihelion measurements of comet 67P, which had a dust/gas ratio of $\sim 4$ \citep{rotundi2015}. However we caution these estimates remain highly uncertain.

\subsection{Nucleus size}
\label{sec:nuc_size}

While it is premature to do any detailed modeling with the current limited data, we can explore the parameter space further to place some limits on the nucleus size from the ground-based photometry and the CN production rate.
First, assuming that all of the CFHT flux within a 5$''$ radius photometry aperture for data obtained is scattered light from a nucleus with an albedo of 0.04, this implies a nucleus radius, $R_N$ $\sim$ 8 km.  However, given that there is visible dust in the coma, this is an extreme upper limit. We can use a surface ice sublimation model \citep{meech1986, meech2004} with constraints on the gas production from our CN observations to investigate the activity for 2I, and get some information about the minimum nucleus radius.  

If we assume that the CN/OH ratio for 2I/Borisov is typical of Solar System comets \citep{ahearn1995}, this implies a water production rate, Q(H$_2$O) = (1.3--5.1)$\times$ 10$^{27}$ molec s$^{-1}$. If, on the other hand we assume that 2I is depleted in CN at the level inferred for 1I  from the amount of outgassing by water needed to explain its non-gravitational acceleration combined with the non-detection of CN \citep{micheli2018}, the suggested production rate is Q(H$_2$O) $\sim$7 $\times$ 10$^{27}$ molec s$^{-1}$. Finally, if 2I had the chemistry of the severely depleted comet 96P/Machholz \citep{schleicher2008}, then the inferred production rate is $\sim$1.1 $\times$ 10$^{29}$ molec s$^{-1}$.

\begin{figure*}[ht!]
\begin{center}
\includegraphics[scale=0.5]{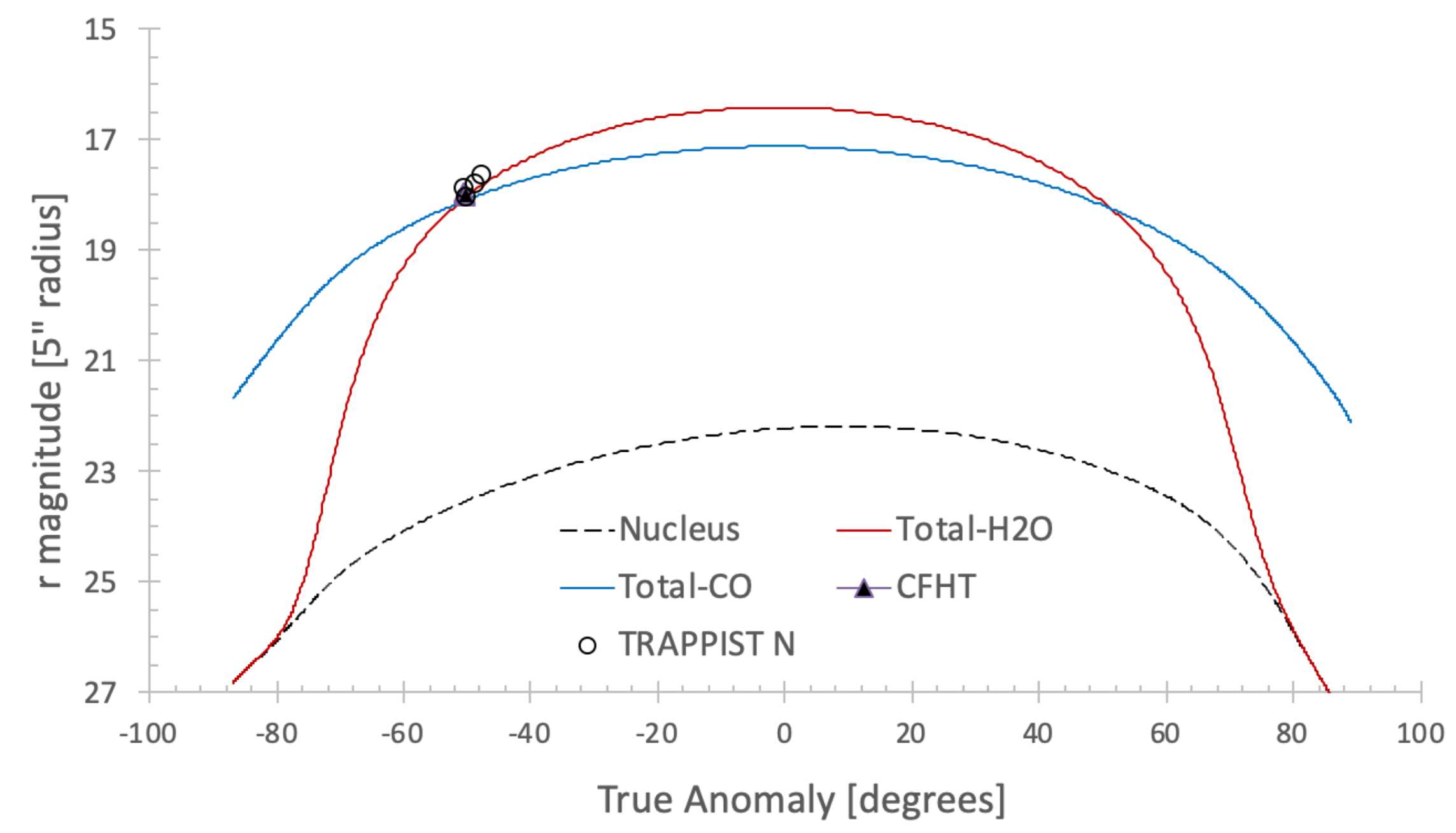}
\caption{Predicted brightness of 2I as a function of true anomaly using our model described in the text, consistent with \added{magnitudes measured from CFHT and TN (closed and open symbols respectively)}. Red and blue curves show predicted brightness for water-dominated and carbon monoxide-dominated sublimation respectively. A true anomaly of $0^\circ$ represents perihelion.}
\label{fig:model}
\end{center}
\end{figure*}

The model computes the amount of gas sublimating from an icy surface exposed to solar heating.  The total brightness within a fixed aperture combines radiation scattered from both the nucleus and the dust dragged from the nucleus in the escaping gas flow, assuming a dust to gas mass ratio of 1. This type of model can distinguish between H$_2$O, CO, and CO$_2$ driven activity.  The model free parameters include: nucleus radius, albedo, emissivity, nucleus density, dust properties, and fractional active area.  

The shape of the light curve -- {\it i.e.} where the curve is steep or shallow -- is determined by the sublimating ice composition.  With reasonable estimates of nucleus size, albedo, density, and grain properties, the fractional active surface area is adjusted to produce the observed volatile production rates. We assume an albedo of 0.04 for both the nucleus and dust and a linear phase function of 0.04 mag deg$^{-1}$ typical of other comets.  We assume a nucleus density of 400 kg m$^{-3}$ similar to that seen for comets 9P/Tempel 1, 103P/Hartley 2 \citep{thomas2009} and 67P/Churyumov-Gerasimenko \citep{patzold2016}, a grain density of 1000 kg m$^{-3}$, and micron-sized grains. Taking a typical fractional active sublimation area of 4\% seen for most comets \citep{ahearn1995}, we can fit the data for a nucleus of radius $R_N$ = 3.3 km, but only for larger grains ($\sim$ 20 $\mu$m). 
Alternatively, a model with a nucleus radius as small as $R_N$ = 0.7 km can fit the data, but only if 100\% of the surface is active, again using large grains (see Fig.~\ref{fig:model}).
The larger grains require more gas to lift, and contribute less to the scattered brightness from the coma. Our model cannot differentiate between this range in radii with
the
present data. However the lack of previous detections of active ISOs would favour a smaller size.

Additionally, we ran a canonical model assuming that 2I was driven by CO sublimation, shown as the blue curve in Fig.~\ref{fig:model}. If this occurred then our model predicts a search for precovery observations of 2I would be worthwhile, albeit noting the small solar phase angle in mid-2019. A pure water ice sublimation model predicts that 2I would have been likely very faint prior to going into solar conjunction around May 2019. A non-detection of the comet in such data would provide significant support for H$_2$O being the main activity driver of 2I.

\section{Conclusions}
\label{sec:conclusions}
We present the first spectroscopic detection of gas emitted by an ISO, 2I/Borisov, on 2019 September 20, using the 4.2 m William Herschel Telescope. Combining these spectra with broad-band photometry 24 hours earlier from the 0.6 m TRAPPIST-North telescope, we present the following results.

\begin{enumerate}
\item 2I/Borisov has CN gas present in its coma, with a gas production rate of Q(CN)$=4\times10^{24}$ s$^{-1}$ at 2.7 au.

\item The upper limit of Q(C$_2$)$\leq 4\times10^{24}$ s$^{-1}$ is consistent with gas abundance ratios measured in Solar System comets.

\item The dust reflectance spectrum is redder at shorter wavelengths, in agreement with previous studies of cometary dust.

\item The true gas/dust ratio is poorly constrained due to the unknown dust size distribution, however the observed Q(CN)/$Af\rho$ is unremarkable when compared to the previously observed population of Solar System comets. The real dust/gas mass ratio is also likely similar to
Solar System
comets.

\item Assuming that the properties of 2I are similar to
Solar System
comets, it implies that the nucleus may be between 0.7-3.3 km, is ejecting large grains, and if it is on the small end, the comet likely has a large fraction of the surface active.

\end{enumerate}

Models of planetary system predict that the formation of the icy planetesimals we call comets should be ubiquitous \citep{raymond2018}. Yet models of protoplanetary disks predict a range of possible disk compositions. For example, the HCN/H$_2$O ratio can vary by a factor $>100$ in protoplanetary disks depending on distance from the star and evolutionary stage 
\citep{eistrup2019}. We also know that within the Solar System, $\sim 30$\% of comets exhibit strong depletions in carbon-chain molecules such as C$_2$ and C$_3$ \citep{ahearn1995}, and significant variations in nuclear ice species such as HCN and CH$_3$OH exist \citep{mumma2011}.
The observation of primary ice species such as H$_2$O, CO$_2$, CO, or their dissociation products, would place the analysis presented here on a much firmer footing.  Currently, our knowledge of the composition of 2I is still relatively
unconstrained. For example,
we do not yet know if 2I is depleted in C$_2$. Yet our data also show it is not C$_2$ rich, and the CN/dust ratio is normal when compared to other comets. The normal scalelengths for production of cometary CN via photodissociation match the observed column density distribution, indicating similar production pathways.
If it were not for its interstellar nature, our current data shows that 2I/Borisov would appear as a rather unremarkable comet in terms of activity and coma composition.

{\it Acknowledgements} We thank the anonymous referee for rapid and helpful comments on the initial version of this manuscript. We also thank Lilian Dominguez and Ian Skillen of the Isaac Newton Group for performing these observations for us at short notice under service programme SW2019b04. The WHT is operated on the island of La Palma by the Isaac Newton Group of Telescopes in the Spanish Observatorio del Roque de los Muchachos of the Instituto de Astrofísica de Canarias. TRAPPIST is a project funded by the Belgian Fonds (National) de la Recherche Scientifique (F.R.S.-FNRS) under grant FRFC 2.5.594.09.F. TRAPPIST-North is a project funded by the University of Liege, in collaboration with Cadi Ayyad University of Marrakech (Morocco). E.J is F.R.S.-FNRS Senior Research Associate. AF and CS acknowledge support for this work from UK STFC grants ST/P0003094/1 and ST/L004569/1. KJM, JTK, and JVK acknowledge support through awards from NASA 80NSSC18K0853.

\bibliography{2I-CN-no-trackchanges}
\bibliographystyle{aasjournal}

\end{document}